\documentclass[prd,twocolumn,superscriptaddress,nofootinbib,preprintnumbers]{revtex4-2}
\usepackage{graphicx} 
\usepackage{amsmath}
\usepackage{amssymb}
\usepackage{hyperref}
\usepackage[caption = false]{subfig}

\usepackage{color}
\definecolor{darkred}{rgb}{1.0,0.1,0.1}
\definecolor{darkgreen}{rgb}{0.1,0.7,0.1}
\definecolor{darkblue}{rgb}{0.1,0.1,1.0}
\definecolor{darkorange}{rgb}{0.9,0.5,0.0}

\newcommand{\Sec}[1]{Sec.~\ref{sec:#1}}

\newcommand{\Reference}[1]{Ref.~\cite{#1}}
\newcommand{\Eq}[1]{Eq.~\eqref{eq:#1}}
\newcommand{\Fig}[1]{Fig.~\ref{fig:#1}}

\begin{document}

\title{The Well-Tempered Likelihood:\\ Honest Confidence Intervals for Misspecified Models}

\preprint{MIT-CTP/6064}

\author{Benjamin~Nachman}
\email{nachman@stanford.edu}
\affiliation{Department of Particle Physics and Astrophysics, Stanford University, Stanford, CA 94305, USA}
\affiliation{Fundamental Physics Directorate, SLAC National Accelerator Laboratory, Menlo Park, CA 94025, USA}

\author{Jesse~Thaler}
\email{jthaler@mit.edu}
\affiliation{Center for Theoretical Physics -- a Leinweber Institute, Massachusetts Institute of Technology, Cambridge, MA 02139, USA}
\affiliation{Institut des Hautes \'Etudes Scientifiques, 91440 Bures-sur-Yvette, France}
\affiliation{Institut de Physique Th\'eorique, CEA Paris-Saclay, 91191 Gif-sur-Yvette, France}
\affiliation{The NSF Institute for Artificial Intelligence and Fundamental Interactions}


\begin{abstract}
Likelihood-based inference in particle physics, and in the physical sciences more broadly, relies on the assumption that the model accurately describes the data.
When the model is misspecified, though, standard confidence intervals shrink to zero width with increasing data, producing overconfident and potentially misleading constraints.
We propose the \emph{well-tempered likelihood}, which divides the likelihood-ratio test statistic by a goodness-of-fit (GOF) statistic evaluated at the best-fit point.
Under correct specification, the GOF is $\mathcal{O}(1)$ and standard inference is recovered.
Under misspecification, both the likelihood and the GOF scale as $\mathcal{O}(N)$ for $N$ data points, so their ratio remains $\mathcal{O}(1)$ and the resulting well-tempered confidence interval self-limits at a floor determined by the model's inadequacy, essentially reducing the effective sample size.
In other words, \textit{all models are correct, as long as your dataset is small enough}.
We present binned and unbinned formulations of the well-tempered likelihood---the latter based on a classifier two-sample test---and demonstrate our method on a Gaussian example and on a measurement of the strong coupling constant using synthetic electron-positron collisions.
\end{abstract}

\maketitle
\tableofcontents

\section{Introduction}

High energy physics is in a precision era where large volumes of complex data are used to infer fundamental parameters of nature with narrowing confidence intervals.
These advances rest on increasingly accurate simulations that take the fundamental parameters as input and forward model the physics and detector effects relevant for a given experiment.
This enables sophisticated strategies for simulation-based inference~\cite{Cranmer:2019eaq}, in which the parameters are identified by finding the synthetic data that best match the observed experimental data.
Despite the success of this program, there is an uncomfortable flaw: the quality of the model fit to data is not taken into account in the inference pipeline.

In the current paradigm, one can obtain narrow confidence intervals on model parameters even if the model does not precisely describe the data.
In practice, bad model fits are hidden by the addition of systematic uncertainties with no statistical origin (SUnSOs).
For example, there may be two programs available for generating some part of the full simulation stack and the difference between programs is reported as an uncertainty.
This uncertainty is then combined with other uncertainties in quadrature as if it were statistical.
More importantly, one or both programs may lead to poor modeling of the data, which is less of an ``uncertainty'' and more of an ``error''~\cite{Barlow:2002yb}.
SUnSOs are more pervasive than widely appreciated, precisely because they can be hidden.%
\footnote{For example, experimental uncertainties like the jet energy scale at colliders are often assumed statistical, but their uncertainties also arise from a parameter estimation measurement with lurking SUnSOs such as from simulating the hadronic recoil in object balancing studies~\cite{ATLAS:2020cli,CMS:2016lmd}.}
Complex analyses using the tools of statistics have been conducted to constrain model parameters, but layering formalism onto a tenuous setup will not yield any new physical insights.

In statistics, the usual solution to a bad fit is to add more parameters to the model.
In fundamental physics, though, we often need to perform inference using imperfect models even though we know they are at best an approximation of nature.
Ideally, we would augment simulations with additional nuisance parameters to capture otherwise mismodeled physical effects, but in practice, it is usually not possible to simply add more parameters to better describe the data.
If the model really is fixed, then the only option to achieve more reliable inference is to (effectively) change the data.%
\footnote{During the completion of  this work, we learned of \Reference{Szewc:2026nvu}, which takes a similar philosophy.  There the impact of an imperfect machine-learned likelihood is mitigated by enforcing a minimum resolution in the analysis, effectively coarsening the data.}

In this paper, we propose the \emph{well-tempered likelihood}, which reduces the effective number of events until the model is a good fit to the data.
The core idea is simple: divide the standard likelihood-ratio test statistic $\Lambda(\theta)$ by the goodness-of-fit (GOF) per degree of freedom $\hat\gamma$, evaluated at the best-fit point, to create a new test statistic:
\begin{equation}
T(\theta) = \frac{\Lambda(\theta)}{\hat{\gamma}}\,.
\label{eq:T_intro}
\end{equation}
Under correct model specification, $\hat\gamma\to 1$ and standard inference is recovered.
Under misspecification, though, both $\Lambda$ and $\hat\gamma$ grow as $\mathcal{O}(N)$ with $N$ data points, so their ratio remains $\mathcal{O}(1)$---the resulting confidence interval has self-limiting rather than collapsing behavior.
The effective sample size,
\begin{equation}
N_{\text{eff}} = \frac{N}{\hat{\gamma}}\,,
\label{eq:Neff_intro}
\end{equation}
saturates at a constant determined by the degree of misspecification and the resolution of the GOF test.
In this way, we can still use the full dataset for inference, but the confidence in poorly specified models is tempered.

The idea of adjusting inference for model misspecification has a long history.
The Particle Data Group (PDG) inflates per-measurement uncertainties by $S=\sqrt{\chi^2/\text{dof}}$ when averaging discrepant measurements with $\chi^2/\text{dof}>1$~\cite{ParticleDataGroup:2024cfk}, which is equivalent to our binned formulation applied to a weighted average, albeit with a different (systematic instead of statistical) interpretation.
The sandwich (Huber--White) estimator~\cite{Huber:1967,White:1982} provides standard errors that are asymptotically correct under misspecification, where its intervals (like the well-tempered ones) are centered near the pseudo-true value $\theta^*$.
The key difference is that sandwich intervals continue to shrink as $1/\sqrt{N}$, converging precisely to the wrong value, whereas the well-tempered interval self-limits at a floor that reflects the model's inadequacy.
Tempered posteriors~\cite{Grunwald:2017,Miller:2019} raise the likelihood to a fractional power $\eta<1$ to safeguard against misspecification; while $\eta$ is often specified \emph{a priori}, data-driven approaches such as SafeBayes~\cite{Grunwald:2017} estimate $\eta$ from the data by minimizing a regret criterion, yielding an effective $\hat\gamma = 1/\eta$.
Our well-tempered approach is distinct in that $\hat\gamma$ is determined entirely by a GOF test rather than a learning rate, providing a self-limiting guarantee without external tuning.

The rest of this paper is organized as follows.
In \Sec{methods}, we review standard likelihood-based inference and introduce the well-tempered test statistic in both binned and unbinned settings.
\Sec{gaussian} presents a Gaussian example that admits exact analytical treatment, and \Sec{physics} applies the method to a particle physics example using unbinned classifier-based inference for the strong coupling constant from jet substructure simulations.
We discuss implications and future directions in \Sec{conclusions}.

\section{Likelihood Tempering by Goodness of Fit}
\label{sec:methods}

\subsection{Review of Maximum Likelihood Inference}

We start with a review of standard statistical inference to set notation.
Consider $N$ observations $X_1,\ldots,X_N$ drawn from an unknown distribution $P$, and a parametric model $f(x;\theta)$ with parameters $\theta\in\mathbb{R}^k$.
The log-likelihood is:
\begin{equation}
\ell(\theta)=\sum_{i=1}^N \log f(X_i;\theta),
\end{equation}
and the maximum likelihood estimator (MLE) for $\theta$ is:
\begin{equation}
\label{eq:MLE}
\hat{\theta}_{\rm MLE}=\arg\max_\theta \ell(\theta).
\end{equation}
The log-likelihood ratio%
\footnote{There are no nuisance parameters in our analysis for simplicity, but here and elsewhere, this could be replaced with the profile likelihood with minimal change to the formalism.}
for testing a parameter value $\theta$ is:
\begin{equation}
\Lambda(\theta) = 2\left(\ell(\hat{\theta}_{\rm MLE}) - \ell(\theta)\right),
\label{eq:profileLR}
\end{equation}
which we will refer to as the ``standard likelihood-ratio test statistic'' (or just ``standard test statistic'' for short).
Under correct model specification, Wilks' theorem~\cite{wilks1938large} guarantees that the standard test statistic converges to a chi-squared distribution with $k$ degrees of freedom:
\begin{equation}
\Lambda(\theta_0)\xrightarrow{d}\chi^2_k,
\end{equation}
where $\theta_0$ is the true parameter value.
A confidence interval at level $1-\alpha$ is then obtained by inverting this test:
\begin{equation}
\{\theta : \Lambda(\theta) \leq \chi^2_{k,\,1-\alpha}\},
\end{equation}
where $\chi^2_{k,\,1-\alpha}$ is the $(1-\alpha)$ quantile of the $\chi^2_k$ distribution.  For a single parameter of interest ($k=1$), this yields intervals that shrink as $1/\sqrt{N}$, reflecting the increasing precision of the MLE with more data.

\subsection{Philosophy Behind Tempering}

When the model is misspecified, the MLE converges not to the true value $\theta_0$ but to the pseudo-true value~\cite{White:1982}:
\begin{equation}
\theta^*=\arg\min_\theta D_{\text{KL}}(P\|f_\theta),
\end{equation}
i.e.~the parameter that minimizes the Kullback--Leibler divergence from the true distribution to the model.
At $\theta^*$, the standard test statistic grows without bound: $\Lambda(\theta^*)=\mathcal{O}(N)$.
If we erroneously assume that Wilks' theorem still holds, the resulting confidence intervals shrink to zero width as $N\to\infty$, producing spuriously precise constraints centered on the wrong value.

The well-tempered likelihood is designed to satisfy two requirements:
\begin{itemize}
\item \textbf{Fidelity.}  Under correct model specification, the well-tempered test statistic should reproduce the standard likelihood-ratio test statistic, so that no statistical power is lost when the model is adequate.
\item \textbf{Self-limitation.}  Under misspecification, the confidence interval should stabilize at a finite floor rather than collapsing to zero, honestly reflecting the model's limitations.
\end{itemize}
We emphasize that while fidelity is already satisfied by standard inference pipelines used in particle physics, self-limitation is a design choice about how to report results when standard inference could be misleading.%
\footnote{An alternative design choice would be to simply refuse to report a confidence interval if the GOF is sufficiently poor.  This is misleading in a different way, since we rarely deal with models so inadequate that the pseudo-true value $\theta^*$ is completely unrelated to the true value $\theta_0$.}

Both of these requirements can be met by the scaling argument previewed in the introduction.
Specifically, consider a GOF test $G(\theta)$, whose value per degree of freedom at the best-fit point is:
\begin{equation}
\label{eq:gamma_def}
\hat{\gamma} \equiv \frac{G(\hat{\theta}_{\rm MLE})}{\text{dof}},
\end{equation}
where the degrees of freedom depend on the specific GOF test being used.
The key property of $\hat\gamma$ is:
\begin{equation}
\hat\gamma = \begin{cases} \mathcal{O}(1) & \text{under correct specification,}\\[4pt] \mathcal{O}(N) & \text{under misspecification.} \end{cases}
\label{eq:GOF_scaling}
\end{equation}
Since $\Lambda(\theta_0)\sim\chi^2_k$ under correct specification, dividing by $\hat\gamma=\mathcal{O}(1)$ preserves standard inference.
Under misspecification, both $\Lambda$ and $\hat\gamma$ grow as $\mathcal{O}(N)$, so $T(\theta)=\Lambda(\theta)/\hat\gamma$ remains $\mathcal{O}(1)$ and the confidence interval self-limits at a constant floor rather than shrinking indefinitely.

One way to characterize this self-limiting property is by the effective sample size $N_{\text{eff}} = N/\hat{\gamma}$ from \Eq{Neff_intro}.
Roughly speaking, $N_{\text{eff}}$ is the number of events below which the GOF measure would not be able to distinguish the data from a misspecified model.
More accurately, $N_{\text{eff}}$ characterizes the rate at which the well-tempered test statistic accumulates information, which saturates when the GOF detects misspecification.
We note that $N_{\text{eff}}$ is best understood as a convenient summary of the well-tempered statistic's scaling behavior, rather than a true effective sample size in the importance-sampling sense.

Any GOF statistic with the scaling property in \Eq{GOF_scaling} can serve as $\hat\gamma$.
In this work, we consider two concrete realizations: one based on the binned Pearson $\chi^2$ statistic, and one based on an unbinned classifier two-sample test.
Inspired by the well-tempered likelihood, we plan to study alternative GOF measures in future work, with the goal of more closely linking GOF testing to parameter inference.

\subsection{Binned Case via Pearson Chi-Squared}
\label{sec:binned_approach}

In the binned setting, suppose the data populate $B$ bins with observed counts $n_b$ and expected counts $\nu_b(\theta)=N\,p_b(\theta)$.  At any parameter value $\theta$, the Pearson chi-squared GOF statistic is:
\begin{equation}
G_{\rm bin}(\theta) = \sum_{b=1}^{B}\frac{\left(n_b - \nu_b(\theta)\right)^2}{\nu_b(\theta)}\,.
\label{eq:pearsonchi2}
\end{equation}
Under correct specification, $G_{\rm bin}(\hat{\theta}_{\rm MLE})\xrightarrow{d}\chi^2_{B-k-1}$ (with $B-k-1$ degrees of freedom after fitting $k$ parameters), while under misspecification $G_{\rm bin}(\theta^*)=\mathcal{O}(N)$.

The well-tempered test statistic is the ratio:
\begin{equation}
T(\theta) = \frac{\Lambda(\theta)}{\hat{\gamma}}\,,\quad \hat{\gamma} = \max\!\left(1,\;\frac{G_{\rm bin}(\hat{\theta}_{\rm MLE})}{B-k-1}\right),
\label{eq:Tbinned}
\end{equation}
where $\hat{\gamma}$ from \Eq{gamma_def} has been clipped below at unity so that the well-tempered interval is never narrower than the standard interval.
Under correct specification, $G_{\rm bin}(\hat{\theta}_{\rm MLE})/(B{-}k{-}1)\to 1$ in probability, the clipping is inactive, and $T(\theta_0)\xrightarrow{d}\chi^2_k$, recovering standard likelihood-ratio inference.
Under misspecification, both $\Lambda(\theta^*)$ and $G_{\rm bin}(\hat{\theta}_{\rm MLE})$ grow as $\mathcal{O}(N)$, so their ratio $T(\theta^*)$ remains $\mathcal{O}(1)$---the confidence intervals \emph{self-limit} rather than collapsing.

More precisely, when $\hat{\gamma}>1$ and under correct specification, $T(\theta_0)/k$ follows an $F_{k,\,B-k-1}$ distribution provided that $\Lambda(\theta_0) \sim \chi^2_k$ and $G_{\rm bin}(\hat{\theta}_{\rm MLE}) \sim \chi^2_{B-k-1}$ are independent.
This follows from the definition of the $F$ distribution as the probability distribution of a ratio of two independent chi-square distributed random variables, divided by their respective degrees of freedom:
\begin{equation}
\frac{T(\theta_0)}{k} = \frac{\Lambda(\theta_0)/k}{G_{\rm bin}(\hat{\theta}_{\rm MLE})/(B-k-1)} \sim F_{k,\,B-k-1}.
\end{equation}
In the Gaussian location family, the independence of $\Lambda(\theta_0)$ and $G_{\rm bin}(\hat{\theta}_{\rm MLE})$ holds exactly by Cochran's theorem~\cite{Cochran1934}, since $\Lambda$ depends only on the sample mean while $G_{\rm bin}(\hat{\mu}_{\rm MLE})$ depends only on the sample deviations.
In the general case, asymptotic independence follows from the orthogonality of the efficient score (which determines $\Lambda$) and the residual GOF components in the asymptotic expansion of the multinomial likelihood~\cite{Ferguson:1996}.
For large $B$, the $F_{k,\,B-k-1}$ distribution approaches $\chi^2_k/k$, but for moderate $B$, its heavier tails provide a natural finite-sample correction.

\subsection{Unbinned Case via Binary Classification}
\label{sec:unbinned_approach}

For unbinned data, there is no uniformly most powerful GOF test~\cite{Williams:2010vh}.
There are a number of possibilities and we focus here on a classifier-based two-sample test~\cite{Friedman:2003,Lopez-Paz:2016} as the GOF statistic, mentioning alternative choices in \Sec{conclusions}.
Suppose that we generate $N$ synthetic samples $\widetilde{X}_1,\ldots,\widetilde{X}_N$ from the fitted model $f(x;\hat{\theta}_{\rm MLE})$, and train a binary classifier $\hat{c}(x)$ to distinguish observed data $\{X_i\}$ (labeled~1) from synthetic samples $\{\widetilde{X}_j\}$ (labeled~0).
In the large-sample limit, the optimal classifier approximates the density ratio~\cite{neyman1933problem}:
\begin{equation}
\frac{\hat{c}(x)}{1-\hat{c}(x)} \;\xrightarrow{\;p\;}\; \frac{p(x)}{f(x;\hat{\theta}_{\rm MLE})}\,,
\label{eq:densityratio}
\end{equation}
where $p(x)$ is the true data-generating density.
Here, we have convergence in probability under standard regularity conditions on the classifier family (universal approximation and sufficient capacity).%
\footnote{Note that this GOF classifier $\hat{c}$ is distinct from the DCTR classifier $\hat{f}$ used to construct the $\theta$-dependent likelihood ratio in \Eq{dctr_llr} below.  Here, $\hat{c}$ is trained solely to test goodness-of-fit at the best-fit point $\hat{\theta}_{\rm MLE}$.}
The classifier-based GOF statistic is the deviance relative to the null hypothesis $\hat{c}_0=\tfrac{1}{2}$ (i.e.~data and model are indistinguishable):
\begin{equation}
G_{\rm cls} = 2\!\left[\sum_{i=1}^{N}\log \frac{\hat{c}(X_i)}{\hat{c}_0} + \sum_{j=1}^{N}\log \frac{1\!-\!\hat{c}(\widetilde{X}_j)}{1-\hat{c}_0}\right].
\label{eq:GC}
\end{equation}
Note that $G_{\rm cls}$ depends implicitly on $\hat{\theta}_{\rm MLE}$, but we omit it as a function argument since this dependence enters only through the synthetic samples $\{\widetilde{X}_j\}$ drawn from $f(x;\hat{\theta}_{\rm MLE})$.
Under correct specification, $p(x)=f(x;\hat{\theta}_{\rm MLE})$ asymptotically, so $\hat{c}\to\hat{c}_0$ and $G_{\rm cls}$ remains $\mathcal{O}(1)$.
Under misspecification, the classifier exploits shape differences between $p(x)$ and $f(x;\hat{\theta}_{\rm MLE})$, such that $G_{\rm cls}=\mathcal{O}(N)$.

The well-tempered statistic in the unbinned case is analogous to \Eq{Tbinned}, though we use a shifted $z$-score rather than a $\chi^2$/dof form for reasons explained later:
\begin{equation}
T(\theta) = \frac{\Lambda(\theta)}{\hat{\gamma}}\,,\quad \hat{\gamma} = \max\!\left(1,\;1 + \frac{G_{\rm cls} - \nu(N)}{\sigma(N)}\right).
\label{eq:Tunbinned}
\end{equation}
Here, $\nu(N)=\langle G_{\rm cls}\rangle_0$ and $\sigma(N)=\mathrm{std}[G_{\rm cls}]_0$ are the mean and standard deviation of $G_{\rm cls}$ under the null hypothesis at sample size $N$.
The quantity $z\equiv(G_{\rm cls}-\nu)/\sigma$ is the significance of the GOF test in units of its null spread.
Under correct specification, $z=\mathcal{O}(1)$ and $\hat{\gamma}\to 1$, recovering $T(\theta_0)\xrightarrow{d}\chi^2_k$.
Under misspecification, $G_{\rm cls}$ acquires an excess that grows as $\mathcal{O}(N)$ while $\sigma$ remains $\mathcal{O}(1)$, so $\hat{\gamma}=\mathcal{O}(N)$ and $T(\theta^*)=\mathcal{O}(1)$, ensuring self-limiting intervals.
As in \Eq{Tbinned}, clipping $\hat{\gamma}$ at 1 is a design choice ensuring that the well-tempered interval is never narrower than the standard interval; it introduces a kink (but not a discontinuity) in $T$ at $z=0$, which in practice has negligible effect because $\hat{\gamma}\approx 1$ in the transition region regardless.

Unlike the binned case where the number of degrees of freedom $B-k-1$ is known analytically, no robust notion of degrees of freedom exists for a classifier-based two-sample test.
In particular, the null distribution of $G_{\rm cls}$ depends on both the classifier architecture and the sample size $N$, so the null parameters $\nu(N)$ and $\sigma(N)$ must be estimated empirically.
This is the reason we chose to use a shifted $z$-score form for $\hat{\gamma}$.
We propose to calibrate these per-$N$ null parameters via permutation: draw $M$ random splits of a reference sample generated at $\hat\theta_{\rm MLE}$, train the GOF classifier on each split, compute $G_{\rm cls}$, and set $\nu(N)=\langle G_{\rm cls}\rangle_0$ and $\sigma(N)=\mathrm{std}[G_{\rm cls}]_0$ to their empirical values over the $M$ splits.
For a finite-capacity classifier under the null, $\hat{c}\approx\tfrac{1}{2}$ with small fluctuations, and each event's contribution to $G_{\rm cls}$ has a negative bias from Jensen's inequality~\cite{jensen1906fonctions}: $\langle\log 2\hat{c}\rangle < \log 2\langle\hat{c}\rangle = 0$.
Empirically, $\nu(N)$ is negative and approximately constant in $N$, and $\sigma(N)$ is positive (by construction) and likewise approximately constant.
The per-$N$ calibration absorbs these biases so that $\hat\gamma\approx 1$ under correct specification regardless of sample size.

This unbinned approach naturally handles high-dimensional and multivariate data without binning or partitioning, and leverages modern machine-learning architectures to detect arbitrary shape discrepancies between data and model.
The principal trade-off relative to the binned case is that the null distribution of $G_{\rm cls}$ must be calibrated by permutation rather than by a closed-form $\chi^2$ expression.
To maximize statistical power in this classifier-based GOF test at moderate $N$, we use $K$-fold cross-validation: the combined data-plus-synthetic sample is partitioned into $K$ folds, a classifier is trained on each set of $K-1$ folds, and the held-out fold provides out-of-fold predictions.
This ensures that every event contributes to the test statistic while avoiding train-on-test bias.
For concreteness, we use $K=5$ for our numerical studies.

\section{Gaussian Proof of Concept}
\label{sec:gaussian}

To demonstrate the well-tempered likelihood in practice, we start with a univariate proof of concept in which the data are drawn from a Gaussian with unknown mean $\mu_0$ and a fixed variance $\sigma_0^2$:
\begin{equation}
X_i \sim \mathcal{N}(\mu_0,\,\sigma_0^2)\,,\quad i=1,\ldots,N\,.
\end{equation}
We assume that the analyst fits a model $f(x;\mu)=\mathcal{N}(\mu,\,1)$ with unit variance.
When $\sigma_0=1$, the model is correctly specified; when $\sigma_0\neq 1$, it is misspecified.%
\footnote{It might seem strange that we fix the model and then study how inference changes as the data varies.
But this is in fact the realistic scenario, where we have to perform inference with a fixed model, and the quality of that inference depends on what nature throws at us for the data.}

For standard inference, the MLE is $\hat{\mu}=\bar{X}$ (the sample mean) regardless of $\sigma_0$, and the standard likelihood-ratio test statistic for testing a value $\mu$ is:
\begin{equation}
\Lambda(\mu) = N(\bar{X}-\mu)^2.
\label{eq:gaussLR}
\end{equation}
Under the model, $\Lambda(\mu_0)\sim\chi^2_1$, giving a $68\%$~confidence-level (CL) interval $\hat{\mu}\pm 1/\sqrt{N}$.
However, when $\sigma_0\neq 1$, the true sampling variance of $\bar{X}$ is $\sigma_0^2/N$, so $\Lambda(\mu_0)$ is actually distributed as $\sigma_0^2\,\chi^2_1$.
If $\sigma_0>1$ the standard interval undercovers, and if $\sigma_0<1$ it overcovers.  Note that under variance misspecification, the mean is still unbiased:  $\theta^*=\theta_0$ since $\bar X\to\mu_0$ regardless of $\sigma_0$.
This Gaussian proof of concept focuses on coverage/width distortion, while \Sec{physics} below illustrates both bias and coverage issues.

\begin{figure*}
\centering
\subfloat[]{\label{fig:gaussian_width}%
  \includegraphics[width=0.47\textwidth]{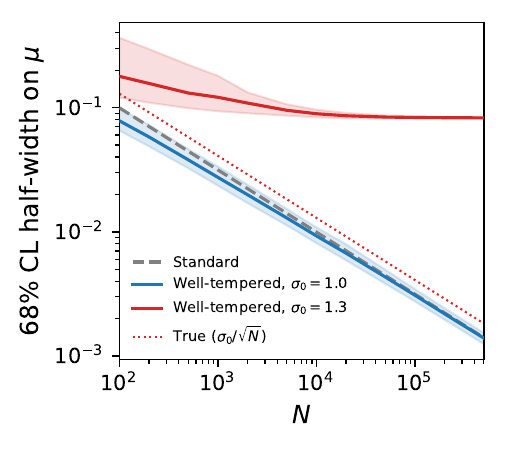}}%
\hfill
\subfloat[]{\label{fig:gaussian_neff}%
  \includegraphics[width=0.47\textwidth]{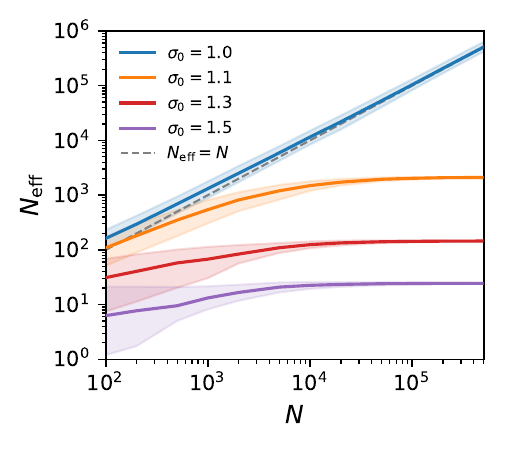}}%
\caption{Gaussian example with the model $\mathcal{N}(\mu,1)$ fit to data from $\mathcal{N}(0,\sigma_0^2)$ using $B=50$ bins.
Shaded bands show 16th--84th percentile ranges over 2000 pseudoexperiments.
\textbf{(a)}~$68\%$~CL interval half-width on $\mu$ versus $N$.
The gray dashed line shows the standard likelihood-ratio interval ($\propto 1/\sqrt{N}$); solid lines show the well-tempered intervals.
Under correct specification ($\sigma_0=1$, blue) both agree.
Under misspecification ($\sigma_0=1.3$, red), the standard interval keeps shrinking with wrong coverage, while the well-tempered interval self-limits to a constant floor.
The red dotted line shows the true frequentist width $\sigma_0/\sqrt{N}$.
\textbf{(b)}~Effective number of events $N_{\text{eff}}=N/\hat{\gamma}$ versus $N$ for several values of $\sigma_0$.
Under correct specification $N_{\text{eff}}=N$; under misspecification $N_{\text{eff}}$ saturates at a constant set by the binning resolution and the degree of misspecification.}
\label{fig:gaussian}
\end{figure*}

\subsection{Binned Well-Tempered Statistic}

To demonstrate the binned approach from \Sec{binned_approach}, we bin the data into $B$ equal-width bins spanning $[\hat{\mu}-5,\,\hat{\mu}+5]$.
Let the expected bin fractions be $p_b^{\text{true}}(\sigma_0)$ for the data and $p_b^{\text{model}}$ for the fitted model $\mathcal{N}(\hat{\mu},1)$.
The Pearson statistic $G_{\rm bin}(\hat{\mu}_{\rm MLE})$ from \Eq{pearsonchi2} compares the observed bin counts against $\mathcal{N}(\hat{\mu}_{\rm MLE},1)$.
Under correct specification with $\sigma_0=1$, $\hat{\gamma}=G_{\rm bin}/(B-2)\to 1$ in probability.
Under misspecification, however:
\begin{equation}
\hat{\gamma}\;\xrightarrow{p}\; 1 + \frac{N\,D_{\chi^2}(\sigma_0)}{B-2}\,,
\label{eq:gaussgamma}
\end{equation}
where the binned $\chi^2$ divergence between the true and model distributions is:%
\footnote{Note that the statistic $G_{\rm bin}$ and the divergence $D_{\chi^2}$ are the sample and population versions of the same Pearson $\chi^2$.  Under misspecification, the noncentral-$\chi^2$ mean gives $\mathbb{E}[G_{\rm bin}(\hat\mu_{\rm MLE})]\simeq(B-2)+N\,D_{\chi^2}(\sigma_0)$, yielding \Eq{gaussgamma}.}
\begin{equation}
D_{\chi^2}(\sigma_0)=\sum_b\frac{(p_b^{\text{true}}(\sigma_0)-p_b^{\text{model}})^2}{p_b^{\text{model}}}\,.
\end{equation}
Because $D_{\chi^2}>0$ when $\sigma_0\neq 1$, $\hat{\gamma}$ grows linearly with $N$, and the GOF detects the misspecification with increasing power.

Using $T=\Lambda/\hat{\gamma}$, the well-tempered confidence interval at level $1-\alpha$ is:
\begin{equation}
\{\mu : T(\mu)\leq\chi^2_{1,1-\alpha}\}\,,
\end{equation}
giving a half-width of:
\begin{equation}
w_{\text{WT}} = \sqrt{\frac{\hat{\gamma}}{N}}\,\sqrt{\chi^2_{1,1-\alpha}} \;\xrightarrow{N\to\infty}\; \sqrt{\frac{D_{\chi^2}(\sigma_0)}{B-2}}\,\sqrt{\chi^2_{1,1-\alpha}}\,.
\label{eq:gaussCI}
\end{equation}
This saturates at a \emph{constant floor} under misspecification, independent of $N$.
In contrast, the standard interval has half-width $w_{\text{std}}=\sqrt{\chi^2_{1,1-\alpha}/N}$, which shrinks to zero regardless of model quality.
This is the self-limiting property in action: the well-tempered interval refuses to narrow below the resolution at which the GOF can distinguish data from model.

Equivalently, the \emph{effective number of events} is:
\begin{equation}
N_{\text{eff}} = \frac{N}{\hat{\gamma}}\;\xrightarrow{N\to\infty}\;\frac{B-2}{D_{\chi^2}(\sigma_0)}\,,
\label{eq:Neff}
\end{equation}
a constant determined by the binning resolution and the degree of misspecification.
When $\sigma_0=1$, $D_{\chi^2}=0$ and $N_{\text{eff}}=N$.
As $|\sigma_0-1|$ increases, $N_{\text{eff}}$ saturates at smaller and smaller values.
When the misspecification is mild ($\sigma_0=1+\varepsilon$ for small $\varepsilon$), $D_{\chi^2}=\mathcal{O}(\varepsilon^2)$, so $N_{\text{eff}}\sim(B-2)/\varepsilon^2$ and the self-limiting floor is reached only at $N\sim 1/\varepsilon^2$.
In the limit $\varepsilon\to 0$ at fixed $N$, standard inference is recovered: the method transitions smoothly between the self-limiting and standard regimes without discontinuity.

\subsection{Coverage Properties}

Now that we have established the desired self-limiting behavior, it is important to characterize what the well-tempered interval covers and when.
Under correct specification ($\sigma_0=1$), the coverage of the $68\%$~CL interval will be $\approx 68\%$ across all $N$, as expected from the $\chi^2_1$ (or $F_{1,B-2}$) null distribution.
Under misspecification ($\sigma_0\neq 1$), the coverage depends on $N$ and the severity of misspecification.

At small $N$, the GOF has limited power ($\hat{\gamma}\approx 1$) so the well-tempered interval essentially coincides with the standard interval.
As $N$ increases past $N\sim (B-2)/D_{\chi^2}$ (the scale at which $\hat{\gamma}$ begins to grow substantially), the well-tempered interval widens relative to the standard interval, and coverage rises above the nominal level.
For $N\gg (B-2)/D_{\chi^2}$, the interval saturates at a constant floor and its coverage approaches unity, since the fixed-width floor eventually exceeds $|\theta^*-\theta_0|$ plus any finite fluctuation.

In this way, the well-tempered method is \emph{conservative} at large $N$: it trades the standard interval's dangerous undercoverage for honest overcoverage that reflects the model's inability to resolve the parameter beyond the GOF-limited precision.%
\footnote{For a very poorly specified model, the well-tempered interval might even undercover if the irreducible bias $|\theta^*-\theta_0|$ is sufficiently large.  Still, the severity of the undercoverage will be reduced compared to the standard interval.}
This conservatism is a feature, not a bug---it is the statistical analogue of reporting ``the model cannot distinguish $\alpha_s$ values differing by less than $\delta$'' rather than reporting a spuriously precise value.
We emphasize that overcoverage does not imply a miscalibration of the intervals, since ``coverage'' is not particularly meaningful when performing inference with a misspecified model.

\subsection{Numerical Illustration}

To demonstrate the self-limiting property of the well-tempered intervals and verify the coverage properties derived above, we perform a Monte Carlo study with fixed $\mu_0=0$ and $B=50$ bins.
We vary $\sigma_0$ and $N$, performing 2000 pseudoexperiments per setting.

In \Fig{gaussian_width}, we show the $68\%$~CL interval half-width as a function of $N$, comparing $\sigma_0=1.0$ and $\sigma_0=1.3$.
The standard interval (dashed) shrinks as $1/\sqrt{N}$ in both cases.
The well-tempered interval (solid) tracks $1/\sqrt{N}$ under correct specification (blue) but saturates to a constant floor ($w_{\text{WT}}\approx 0.084$) under misspecification (red), which is precisely the self-limiting behavior we desire.

For comparison, the dotted line shows the true frequentist width $\sigma_0/\sqrt{N}$, which shrinks to zero.
As desired, the well-tempered method is conservative under misspecification, providing wider-than-necessary intervals rather than dangerously narrow ones.
The measured coverage matches the analytical expectations of the previous subsection.
Quantitatively, for $\sigma_0=1.3$, the standard interval covers the true value only $\approx 56\%$ of the time, an undercoverage inherited by the well-tempered interval at small $N$.
The true coverage of the well-tempered interval then exceeds the nominal $68\%$ once $\hat{\gamma}>\sigma_0^2$, which occurs by $N\sim\mathcal{O}(100)$.

In \Fig{gaussian_neff}, we show the number of effective events $N_{\text{eff}}$ versus $N$.
We see the expected diagonal behavior when $\sigma_0=1$, while the curves bend over to a plateau for $\sigma_0\neq 1$.
For $\sigma_0 \in \{1.1,\,1.3,\,1.5\}$, the asymptotic effective sample sizes are $N_{\text{eff}}\approx \{1900, 140, 24\}$.
This is again the desired behavior, where the number of effective events is scaled to match the adequacy of the model.

\section{Physics Inference: Strong Coupling Constant}
\label{sec:physics}

We now demonstrate the well-tempered likelihood on a realistic particle physics example using simulated jet substructure data, with the goal of inferring the strong coupling constant $\alpha_s$.
Like in \Sec{gaussian}, this is a one-dimensional inference problem, but one that involves enough model misspecification to lead to bias.

\subsection{Simulation and Observables}

For this study, particle-level $e^+e^-\!\to Z\to$~hadrons events (i.e.~hadronic $Z$ decays at $\sqrt{s}=91.2$~GeV) are generated with \textsc{Pythia}~8.317~\cite{Bierlich:2022pfr} starting from the Monash tune~\cite{Skands:2014pea} (\texttt{Tune:ee~=~7}).
Final-state particles are clustered with the anti-$k_T$ algorithm~\cite{Cacciari:2008gp} using the \texttt{fastjet} Python package~\cite{Roy:2022rlt} that wraps FastJet~3.5.1~\cite{Cacciari:2011ma}.
We use the $e^+e^-$ anti-$k_T$ measure (i.e.~\texttt{ee\_genkt} with $p=-1$), with a jet radius of $R=0.8$ and E-scheme recombination, and we process the hardest two jets by energy.
Five jet substructure observables define the feature space $x$:
\begin{itemize}
\item Jet mass (from the E-scheme four-vector sum);
\item Constituent multiplicity (number of charged plus neutral objects);
\item Number of charged kaons;
\item Energy dispersion $E^D = \sqrt{\sum_i E_i^2}\,/\,\sum_j E_j$ (the $e^+e^-$ analogue of $p_T^D$~\cite{CMS:2013kfa}); and
\item Jet width $\sum_i E_i\,\Omega_i\,/\,\sum_j E_j$~\cite{Gras:2017jty}, where $\Omega_i$ is the 3D opening angle of constituent $i$ to the E-scheme jet axis.
\end{itemize}
Using energy weights and 3D opening angles matches the $e^+e^-$ context, in contrast to the transverse-momentum weights and rapidity-azimuth distances typically used at hadron colliders.

We generate the \textsc{Pythia} samples ourselves rather than relying on external datasets, which gives us full control over the parameter values, the reference point, and the degree of misspecification.
\textsc{Pythia} has three tuning parameters that we will adjust, with all other parameters held at their default (Monash) values:
\begin{itemize}
\item $\alpha_s$ (\texttt{TimeShower:alphaSvalue}):  strong coupling constant;
\item $a_\text{Lund}$ (\texttt{StringZ:aLund}):  Lund fragmentation parameter;
\item $\text{probStoUD}$ (\texttt{StringFlav:probStoUD}):  strangeness suppression factor.
\end{itemize}
The default values of these parameters are:
\begin{equation}
\label{eq:dctr_reference_values}
(\alpha_s,\,a_\text{Lund},\,\text{probStoUD})=(0.1365,\,0.68,\,0.217).
\end{equation}

For inference, we fit a single parameter $\alpha_s$ while fixing the nuisance parameters $a_\text{Lund}$ and $\text{probStoUD}$ at their default values.
In practice, the fitted model is represented by a neural network surrogate as described in the next subsection.
We fit $\alpha_s$ with this model to two different datasets to illustrate the effect of misspecification:
\begin{itemize}
\item \textbf{Correctly specified:}
The measured ``data'' are generated at $(\alpha_s,\,a_\text{Lund},\,\text{probStoUD})=(0.16,\,0.68,\,0.217)$.
Here, the nuisance values used to generate the data agree with the ones used in the fitted model for inference, so the model is well-specified.
\item \textbf{Misspecified:}
The measured ``data'' are generated at $(\alpha_s,\,a_\text{Lund},\,\text{probStoUD})=(0.16,\,0.40,\,0.10)$.
While the true $\alpha_s=0.16$ is identical to the correctly specified case, the nuisance parameters differ from the assumed values, so the model is misspecified.
This isolates the effect of nuisance misspecification at a fixed true value of the parameter of interest.
\end{itemize}

\subsection{Surrogate Model and Standard Inference}

Before implementing the well-tempered likelihood, we first need to define the standard test statistic $\Lambda(\alpha_s)$.
For this, we use the DCTR (Deep neural networks using Classification for Tuning and Reweighting) approach~\cite{Andreassen:2019nnm}, which uses parametrized classifiers~\cite{Cranmer:2015bka,Baldi:2016fzo} to create a surrogate likelihood ratio.

DCTR learns a continuous per-event reweighting function $\hat{f}(x;\theta)$, trained to distinguish between events from a reference sample and events generated at varying parameter values $\theta$.
For our reference sample, we generate $10^6$ \textsc{Pythia} jets, using the default values from \Eq{dctr_reference_values}.
For our variable-parameter sample, we generate another $10^6$ \textsc{Pythia} jets, where the three tuning parameters vary simultaneously and uniformly over the ranges $\alpha_s\in[0.10,0.18]$, $a_\text{Lund}\in[0.50,0.90]$, and $\text{probStoUD}\in[0.10,0.30]$.

The DCTR reweighting function $\hat{f}(x;\theta)$ is implemented as a feed-forward neural network with residual connections~\cite{He:2015wrn}, hidden layers of widths $(256,256,256,128)$, and layer normalization~\cite{Ba:2016jcy}.
GELU (Gaussian error linear unit) activation functions~\cite{Hendrycks:2016qxa} are used to ensure that $\hat{f}(x;\theta)$ is twice differentiable (see \Eq{std_error} below), and dropout~\cite{Srivastava:2014kpo} with a rate of $0.05$ is used for regularization.
The network inputs are the five observables $x$ and three parameters $\theta$, while the output is the classifier score in $[0,1]$.
Using a training/validation split of 85\%/15\%, we optimize the binary cross-entropy loss using the AdamW optimizer~\cite{Kingma:2014vow,Loshchilov:2017bsp} with a learning rate of $3\times 10^{-4}$ and a weight decay of $10^{-4}$, with early stopping on the held-out validation set (patience of 25 epochs).

After training, the surrogate log-likelihood ratio at parameter $\theta$ relative to the reference $\theta_0$ is:
\begin{equation}
\ell(\theta) - \ell(\theta_0) = \sum_{i=1}^{N}\log\frac{\hat{f}(X_i;\theta)}{1-\hat{f}(X_i;\theta)},
\label{eq:dctr_llr}
\end{equation}
where the data points $X_i$ come from the measured data (not the training data).
To convert \Eq{dctr_llr} into $\Lambda(\alpha_s)$, we simply plug it in to the standard test statistic formula from \Eq{profileLR}, where $\ell(\theta_0)$ cancels in the difference.
Since we are only interested in inferring $\alpha_s$, we fix $(a_\text{Lund},\,\text{probStoUD})$ to their default values when evaluating the DCTR model.
Strictly speaking, training the DCTR model over all three parameters was overkill for this one-parameter study, but we anticipate using this three-parameter model for future studies of profiling.%
\footnote{In the misspecified case, $a_\text{Lund}=0.40$ is below the training range $[0.50,0.90]$ and $\text{probStoUD}=0.10$ is at the lower edge of $[0.10,0.30]$; this is an interesting situation where profiling alone would not result in a well-specified model.  For the current study, the DCTR model is only evaluated at the in-range default nuisance values, so no extrapolation is required.}

Given $\Lambda(\alpha_s)$, we can perform standard inference on $\alpha_s$.
Concretely, we obtain the MLE $\hat{\alpha}_s$ from \Eq{MLE} with the L-BFGS-B algorithm~\cite{Byrd:1995,Zhu:1997} (typically ${\sim}10$ function evaluations per fit).
This minimization is implemented in SciPy~\cite{Virtanen:2019joe}, using exact gradients from automatic differentiation through the network via PyTorch~\cite{Paszke:2019xhz}.
To extract the $68\%$~CL half-width on $\hat{\alpha}_s$, we compute the curvature of the standard test statistic at the minimum using second-order automatic differentiation:
\begin{equation}
\label{eq:std_error}
w_{\rm std}=\frac{1}{\sqrt{H_{\rm std}}}, \qquad H_{\rm std} = \frac{1}{2} \frac{\mathrm{d}^2 \Lambda}{\mathrm{d}\alpha_s^2}.
\end{equation}
For visualization purposes, we find it convenient to rescale the $68\%$~CL intervals to $95\%$~CL intervals.

\begin{figure*}
\centering
\subfloat[]{\label{fig:physics_width_correct}%
  \includegraphics[width=0.32\textwidth]{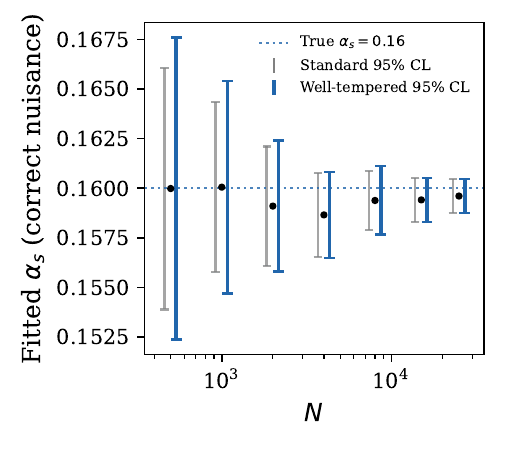}}%
\hfill
\subfloat[]{\label{fig:physics_width}%
  \includegraphics[width=0.32\textwidth]{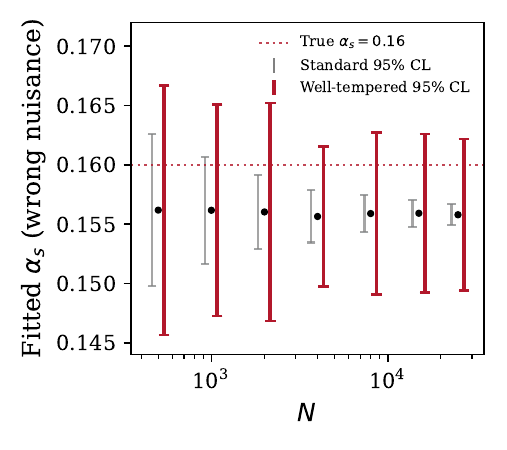}}%
\hfill
\subfloat[]{\label{fig:physics_neff}%
  \includegraphics[width=0.32\textwidth]{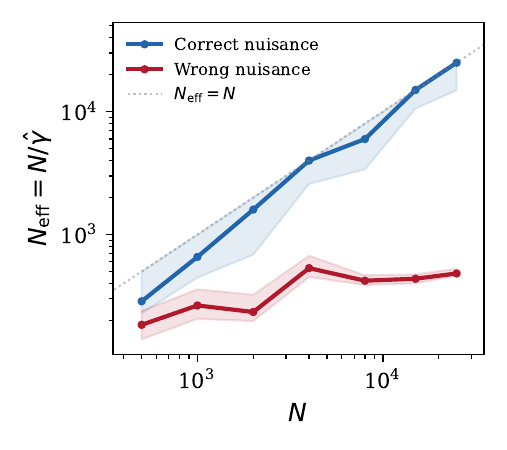}}%
\caption{
Particle physics example using DCTR-based likelihood inference for $\alpha_s$ on \textsc{Pythia}~8 jet substructure data with $e^+e^-$ anti-$k_T$ clustering and energy-based observables (median over 25 pseudoexperiments per point).  \textbf{(a)}~Fitted $\alpha_s$ versus $N$ under correct specification.  The median MLE ($\hat{\alpha}_s\approx 0.160$) is close to the true value $\alpha_s=0.16$ (blue dotted line).  Gray and blue error bars show the standard and well-tempered $95\%$~CL intervals, which are nearly identical when the model is correctly specified.  \textbf{(b)}~Same as (a) but under misspecification (wrong nuisance parameters, same true $\alpha_s=0.16$).  The median MLE ($\hat{\alpha}_s\approx 0.156$) is biased away from the true value $\alpha_s=0.16$ (red dotted line).  The standard interval (gray) shrinks as $1/\sqrt{N}$ and becomes spuriously precise, while the well-tempered interval (red) self-limits.  \textbf{(c)}~$N_\text{eff}=N/\hat{\gamma}$ versus~$N$.  Under correct specification (blue), $N_\text{eff}$ tracks $N$; under misspecification (red), $N_\text{eff}$ self-limits, reaching $N_\text{eff}\approx480$ at $N=25\,000$ as the classifier-based GOF detects the irreducible model--data discrepancy.}
\label{fig:physics}
\end{figure*}

\subsection{Unbinned Well-Tempered Statistic}

The well-tempered test statistic $T(\alpha_s) = \Lambda(\alpha_s)/\hat{\gamma}$ is based on two different machine-learning models that are important to distinguish.
In the previous subsection, we already showed how to use DCTR to build the standard test statistic $\Lambda(\alpha_s)$.
For the GOF statistic $\hat{\gamma}$, we use the unbinned classifier-based approach from \Sec{unbinned_approach}.
Both $\Lambda(\alpha_s)$ and $\hat{\gamma}$ involve training classifiers, but $\Lambda(\alpha_s)$ uses a parametrized classifier dependent on $\alpha_s$, whereas $\hat{\gamma}$ uses a simple two-sample classifier.

To build $\hat{\gamma}$, we use a gradient-boosted decision tree~\cite{Friedman:2001wbq,Chen:2016btl} $\hat{c}(x)$ instead of a neural network.
This classifier is trained to distinguish the $N$ measured data events from $N$ synthetic events.
These $N$ synthetic events are obtained by importance-resampling a 90k-event reference pool with replacement, with probabilities proportional to the DCTR weights at the MLE $\hat\alpha_s$, yielding unit-weight events.
From the classifier $\hat{c}(x)$, we construct the GOF statistic $G_{\rm cls}$ from \Eq{GC}.
As discussed at the end of \Sec{unbinned_approach}, we use 5-fold cross-validation, such that no term in \Eq{GC} is ever evaluated on a data point used during training.
The choice of $K=5$ folds is standard in the cross-validation literature~\cite{Hastie:2009}; we have verified that the results are insensitive to $K$ in the range $3$--$10$.

To convert $G_{\rm cls}$ into $\hat{\gamma}$ from \Eq{Tunbinned}, we need the null parameters $\nu(N)$ and $\sigma(N)$.
These are calibrated per-$N$ via $M=20$ permutations of the reference sample, each involving $K=5$-fold classifier training.
In terms of computational cost, this adds approximately $20\times 5=100$ classifier fits on top of the DCTR likelihood construction (which is one neural network, trained once).
For analyses with fast-to-evaluate models (as in this example, where gradient-boosted trees train in seconds), this overhead is modest, and our full analysis completes in minutes.
For computationally expensive simulations, though, amortized calibration strategies (e.g., interpolating $\nu(N)$ and $\sigma(N)$ from a sparse grid of $N$ values) could substantially reduce the computational cost.

With the full well-tempered test statistic $T(\alpha_s) = \Lambda(\alpha_s)/\hat{\gamma}$ in hand, we perform inference on $\alpha_s$.
Since $\hat{\gamma}$ is independent of $\alpha_s$, the only impact of using the well-tempered test statistic compared to the standard one is a widening of the interval by $\sqrt{\hat{\gamma}}$:
\begin{equation}
w_{\rm WT} = w_{\rm std} \sqrt{\hat{\gamma}}, 
\end{equation}
with the standard width given in \Eq{std_error}.
In the case of a well-specified model, $\hat{\gamma} \approx 1$ and we recover standard confidence intervals.
With a misspecified model, the confidence intervals widen when using the well-tempered test statistic, but the MLE $\hat{\alpha}_s$ is unchanged.

\subsection{Results}

In \Fig{physics}, we show the results of the unbinned well-tempered procedure as a function of $N$ (the number of data events), with medians over 25 pseudoexperiments per point.
The fitted $\alpha_s$ under the correctly-specified scenario is shown in \Fig{physics_width_correct}, with standard (gray) and well-tempered (blue) $95\%$~CL error bars.
In this case, the median MLE $\hat{\alpha}_s\approx 0.160$ is close to the true value $\alpha_s=0.16$, and the standard and well-tempered $95\%$~CL intervals are nearly identical, both shrinking with the desired $1/\sqrt{N}$ scaling.
For the misspecified model in \Fig{physics_width}, the median MLE $\hat{\alpha}_s\approx 0.156$ is biased away from the true value $\alpha_s = 0.16$ due to the wrong nuisance parameters.
The standard interval shrinks as $1/\sqrt{N}$, becoming spuriously precise, while the well-tempered interval (now in red) self-limits and continues to encompass the true value.

We show $N_\text{eff}=N/\hat{\gamma}$ in \Fig{physics_neff}.
Under correct specification, $\hat{\gamma}\approx 1$ and $N_\text{eff}$ tracks $N$ across the full range.
(At small $N$, $N_\text{eff}$ may dip slightly below $N$ due to finite-sample bias in the permutation-based calibration of $\nu(N)$; this artifact diminishes with increasing $N$.)
Under misspecification, $\hat{\gamma}$ grows from ${\approx}\,3$ at $N=500$ to ${\approx}\,52$ at $N=25\,000$, while $N_\text{eff}$ rises only slowly, from $N_\text{eff}\approx180$ at $N=500$ to $\approx480$ at $N=25\,000$---far below $N$ and growing far more slowly than linearly.
In this way, the irreducible model--data discrepancy imposes a GOF penalty that grows with $N$, preventing $N_\text{eff}$ from exceeding the resolution at which the classifier can distinguish data from model.
This is the desired self-limiting property: the well-tempered interval refuses to narrow below the floor set by the model's inadequacy, regardless of how much data is collected.

\section{Conclusions and Outlook}
\label{sec:conclusions}

In this paper, we introduced the well-tempered likelihood, a modification of likelihood-based inference in which the standard test statistic $\Lambda(\theta)$ is divided by a GOF statistic $\hat\gamma$ evaluated at the best-fit point.
Our well-tempered construction ensures two key properties.
First, under correct model specification, $\hat\gamma\to 1$ and standard Wilks-theorem inference is recovered, such that no statistical power is lost when the model is adequate.
Second, under misspecification, $\hat\gamma$ grows as $\mathcal{O}(N)$ in lockstep with $\Lambda$, so the well-tempered statistic remains $\mathcal{O}(1)$ and the confidence interval self-limits at a constant floor rather than collapsing to zero.
The effective sample size $N_{\text{eff}}=N/\hat\gamma$ saturates at a value determined by the resolution of the GOF test and the severity of the misspecification, honestly informing the analyst of the model's limitations.

The well-tempered framework can accommodate any GOF statistic, as long as its null distribution is known or can be calibrated.
In a Gaussian proof-of-concept study, we used a binned GOF statistic, where the binned well-tempered statistic follows a known $F$-distribution.
In a more realistic particle physics study involving jet substructure, we combined the DCTR method for unbinned likelihood construction with a classifier-based two-sample GOF statistic; here, the primary computational cost came from calibrating the associated null parameters of the GOF.
Both studies showcased the self-limiting behavior of the well-tempered interval, where standard inference would yield arbitrarily narrow intervals with incorrect coverage (and, in the physics example, centered on a biased value).

The well-tempered approach has several attractive features for modern particle physics analyses.
It provides a principled, data-driven alternative to SUnSOs: rather than inflating uncertainties by hand, the data themselves reveal when and by how much the model fails.
The method is agnostic to the source of misspecification---whether from missing higher-order corrections, imperfect parton showering, or unmodeled detector effects---because the GOF test detects any discrepancy between data and model.
The unbinned formulation, leveraging classifiers as GOF tests, is particularly well-suited to the high-dimensional, multivariate settings that are increasingly common in LHC analyses and beyond.
Beyond particle physics, our framework applies to any likelihood-based inference problem where model misspecification is a concern; examples in the physical sciences include cosmological parameter estimation, gravitational-wave source characterization, and climate model calibration.
In this way, the well-tempered likelihood provides a general-purpose safeguard against overconfident constraints.

Several extensions of the well-tempered approach merit further investigation.
Our current formulation treats all parameters symmetrically; a natural generalization would be to profile nuisance parameters within the GOF, evaluating the classifier-based test between data and the model at the jointly profiled point $(\hat{\theta}_{\rm MLE},\hat{\eta})$, where $\eta$ denotes nuisance parameters.
This would penalize only the residual misspecification that cannot be absorbed by the model's systematic degrees of freedom, avoiding over-penalization when nuisance parameters legitimately absorb part of the model--data discrepancy.
Such an extension changes the method's behavior qualitatively (e.g.~the saturation level of $N_{\text{eff}}$ would depend on the residual rather than total misspecification) and warrants a dedicated study with realistic nuisance-parameter examples.

Another avenue of investigation is the choice of GOF test, which affects the sensitivity and the saturation level of $N_{\text{eff}}$.
When using a classifier two-sample test, the results depend on the specific architecture, training procedure, and calibration procedure.
A more powerful test (e.g., a larger neural network with greater capacity) detects misspecification at smaller $N$, yielding a lower $N_{\text{eff}}$ plateau and wider self-limiting intervals.
Conversely, a weak or underpowered test is more permissive, allowing narrower intervals.
Potential correlations between errors in the GOF classifier and the DCTR likelihood classifier are a related practical consideration that merits investigation.
Beyond classifier-based approaches, there are other strategies to build two-sample tests, including the kernel maximum mean discrepancy~\cite{Gretton:2012} and energy statistics~\cite{Szekely:2013}; these alternatives slot directly into the $z$-score construction of \Eq{Tunbinned}, since only the null calibration of the GOF statistic changes.

In a frequentist sense, more powerful GOF tests are more conservative (i.e.~safer), since they provide stronger protection against overconfidence.
That said, the analyst should choose a GOF test whose power matches the types of misspecification they consider relevant---this is a modeling choice analogous to selecting a test statistic in hypothesis testing, not a tunable free parameter.
In future work, we plan to develop GOF tests inspired by the task of parameter estimation, where the GOF test prioritizes identifying mismodeling that distorts inferred parameter values, rather than generic mismodeling effects.
Finally, the well-tempered framework could be extended to hypothesis testing, where the question shifts from parameter estimation to model comparison, and overconfident exclusions from misspecified models are equally dangerous.

\emph{Code and data availability.}---All code and data needed to reproduce the results of this paper are available at \url{https://github.com/bnachman/well-tempered-likelihood} and \url{https://zenodo.org/records/21481036}. The repository contains the \textsc{Pythia}~8.317 event-generation scripts, the DCTR training code and the trained network, the analysis and plotting scripts for both the Gaussian and physics examples, and the cached results and test datasets required to regenerate every figure.  The training dataset is on Zenodo.

\begin{acknowledgments}
BN is supported by the Department of Energy, Office of Science under contract number DE-AC02-76SF00515.
JT is supported by the U.S.\ National Science Foundation (NSF) under Cooperative Agreement PHY-2019786 (The NSF AI Institute for Artificial Intelligence and Fundamental Interactions, \url{http://iaifi.org/}), by the U.S.\ Department of Energy (DOE) Office of High Energy Physics under grant number DE-SC0012567, and by the Simons Foundation through Investigator grant 929241.
JT also thanks the Institut des Hautes \'Etudes Scientifiques (IHES) and the Institut de Physique Th\'eorique (IPhT) for providing an inspiring sabbatical environment to carry out this research.
We acknowledge the use of Claude Code (Anthropic) throughout this project---for brainstorming and refining the central idea, for collaborating on the theoretical results, for assistance in scripting and running the numerical studies, and for help editing the manuscript---and we take full responsibility for the content of this manuscript.

\end{acknowledgments}

\bibliography{welltempered_references}

\end{document}